\documentstyle[aps,prd,epsf,floats]{revtex} 

\draft

\begin{document}
\twocolumn[\hsize\textwidth\columnwidth\hsize\csname
@twocolumnfalse\endcsname

\title{Possible Constraints on the Time Variation of the Fine Structure
Constant from Cosmic Microwave Background Data}

\author{Steen Hannestad}

\address{Institute of Physics and Astronomy,
University of Aarhus,
DK-8000 \AA rhus C, Denmark}

\date{\today}

\maketitle

\begin{abstract}
The formation of the cosmic microwave background radiation (CMBR) 
provides a very powerful probe of the early universe at the epoch
of recombination. Specifically, it is possible to constrain the
variation of fundamental physical constants in the early universe.
We have calculated the effect of a varying electromagnetic coupling
constant ($\alpha$) on the CMBR and find that new satellite
experiments should provide a tight constraint on the value of $\alpha$
at recombination which is complementary to existing constraints. 
An estimate of the obtainable precision is $|\dot{\alpha}/\alpha| \leq 
7 \times 10^{-13}\, {\rm y}^{-1}$ in a realistic experiment.
\end{abstract}

\vskip1.8pc]

\section{introduction}

A fundamental question in physics is whether or not the physical
constants are actually constant. Some unifying physical theories 
such as superstring theories do in fact suggest that the physical
``fine structure'' constants change in time \cite{superstring}.
It is therefore of considerable importance to find methods of 
detecting a possible time evolution these quantities.
In the present paper we wish to specifically discuss the electromagnetic
fine structure constant $\alpha$. Its present value is known quite
precisely to be \cite{PDG98}
\begin{equation}
\alpha_0^{-1} = (e^2/4\pi)^{-1} = 1/137.0359895(61).
\end{equation}
One option for detecting time variation is of
course to measure its value in the laboratory and constrain its
time derivative in this way. However, this has the major drawback that
even though quite minute changes are detectable the time differences
are so small that only a moderate sized time derivative is detectable.

Therefore one often turns to other methods. For instance 
it is possible to use
astrophysical arguments to constrain the evolution of $\alpha$,
the most commonly used method being to use differential changes 
in quasar absorption lines.
This method offers both a long look back time (for $z \simeq 3$ one
has $t/t_0 \simeq 1/8$, assuming a standard flat cold dark matter
(CDM) cosmology) and the
ability to detect rather small changes in $\alpha$. Another
possibility is to use Big Bang nucleosynthesis, but this method 
suffers from the problem that constraints on $\alpha$ are based on
a specific assumption on how the neutron to proton mass difference 
depends on $\alpha$.

As a possible probe that is complementary to all the others discussed we 
investigate the sensitivity of the CMBR to changes in $\alpha$. It is well
known that the fluctuation spectrum of the CMBR is extremely sensitive
to the physical conditions at recombination \cite{tegmark}
and, using inversion technique,
it should therefore be possible to determine the physical parameters
at recombination given sufficiently good observations.

The fluctuations are usually described in terms of spherical harmonics
\begin{equation}
T(\theta,\phi) =\sum_{lm}a_{lm}Y_{lm}(\theta,\phi),
\end{equation}
where the coefficients are related to $C_l$ coefficients by
\begin{equation}
C_l \equiv \langle|a_{lm}|^2\rangle.
\end{equation}
These fluctuations were first detected in 1992 by the COBE satellite
\cite{cobe},
but only for $l \lesssim 20$. At such low $l$ the power spectrum is
almost degenerate in the cosmological parameters and no real constraints
are obtainable. In the next few years, however, the power spectrum
will be measured out to $l \simeq 2500$ by two new probes, MAP and
PLANCK \cite{MAP+PLANCK}, and
using this data should yield precision measurements of the
physical parameters at recombination.
It should also be possible to constrain new exotic physics such as 
non-standard neutrinos \cite{hannestad,HET98} or, indeed, 
a change in $\alpha$.

In the next sections we discuss the physical consequences of changing
$\alpha$ and calculate an estimate of how precisely we can hope to
measure such a change with the CMBR data.
To calculate actual CMBR power spectra we have used the CMBFAST
package developed by Seljak and Zaldarriagga \cite{SZ96}.
Finally, as will be discussed later there is also a terrestrial method
which offers long look back times, namely to use the Oklo natural
fission reactor in Gabon. This method offers the currently most
stringent limit on time variation in $\alpha$.

\section{consequences of changing $\alpha$}

Since the formation of the CMBR is based entirely on electromagnetic
processes, changing the strength of these interactions is bound
to change the CMBR fluctuation spectrum.
First of all, it changes the Thomson scattering cross
section for all interacting particles. Second, it also changes the
recombination of hydrogen. The second effect is far more subtle than
the first since it also involves changing all the energy levels of the
hydrogen (and helium) atom. In the following, we shall neglect the
impact on helium and only concentrate on hydrogen.
Notice that there is also a small secondary effect from the change
in helium abundance from nucleosynthesis which we shall also neglect in
the present paper.

{\it Thomson scattering --}
By far the most efficient equilibration mechanism for thermalising the
photon gas in the early universe is Thomson scattering on free electrons
(not protons since the rate for this process is suppressed by a factor
$m_e^2/m_p^2 \simeq 3 \times 10^{-7}$). The fundamental cross section for
this process is given by \cite{weinberg}
\begin{equation}
\sigma_T = \frac{1}{6 \pi} \frac{e^4}{m_e^2},
\end{equation}
meaning that it has a $\alpha^2$ dependence.

{\it Recombination --}
The phenomenon of recombination is of paramount importance for the 
formation of the CMBR since photon equilibration is mediated by
Thomson scattering on free electrons. 
Prior to recombination the photons are tightly coupled to the 
electron-baryon fluid, whereas subsequent to recombination 
the photons are therefore essentially free particles.
Thus the epoch of CMBR formation is directly linked to the recombination
epoch. The recombination of hydrogen has been extensively studied 
by many authors and we shall follow the treatment by Ma and Bertschinger
\cite{MB95} which is based on the earlier treatment by Peebles \cite{peebles}.

Recombination directly to the ground state is strongly prohibited in
the early universe since it leads to immediate re-ionisation. Instead
it proceeds via 2-photon emission from the $2s$ level or via the
redshift of Ly-$\alpha$ photons out of the line center \cite{peebles}. 
Putting together these two effects with the appropriate recombination 
coefficients to all exited levels, one obtains an equation for the
time-evolution of the ionisation fraction, $x_e \equiv n_e/n_H$,
with respect to conformal time
\begin{equation}
\frac{dx_e}{d\tau} = a C_r 
\left[\beta(T_b)(1-x_e)-n_H \alpha^{(2)}(T_b)x_e^2\right],
\end{equation}
where the first term on the right hand side describes collisional 
ionisation from the ground state and the second describes the recombination
rate. $a$ is the cosmological scale factor, normalised so as to be 
equal to one at present, $n_H$ is the total number density of
hydrogen nuclei and $T_b$ is the baryon temperature.
The other factors are given in the following way
\begin{eqnarray}
\alpha^{(2)}(T_b) & = & \frac{64 \pi}{(27 \pi)^{1/2}} 
\frac{e^4}{m_e^2} \left(\frac{T_b}{B_1}\right) \phi_2(T_b) \\
\phi_2(T_b) & \simeq & 0.448 \log \left( \frac{B_1}{T_b}\right) \\
\beta(T) & = & \left(\frac{m_e T_b}{2 \pi}\right)^{3/2} 
e^{-B_1/T_b} \alpha^{(2)}(T_b) \\
B_1 & = & m_e e^2/2  = 13.6 \, {\rm eV}.
\end{eqnarray}
The reduction factor $C_r$ has been calculated by Peebles \cite{peebles}
and is given by
\begin{equation}
C_r = \frac{\Lambda_\alpha + \Lambda_{{\rm 2s} \rightarrow {\rm 1s}}}
{\Lambda_\alpha + \Lambda_{{\rm 2s} \rightarrow {\rm 1s}} + 
\beta^{(2)}(T_b)},
\end{equation}
where
\begin{eqnarray}
\beta^{(2)}(T_b) & = & \beta (T_b) e^{\omega_\alpha/T_b} \\
\Lambda_\alpha & = & \frac{8 \pi \dot{a}}{a^2 \lambda_\alpha^3 (1-x_e) n_H} \\
\lambda_\alpha & = & \frac{8 \pi}{3 B_1}. 
\end{eqnarray}
All these equations scale quite straightforwardly with $\alpha$. The only
thing left is to treat the two-photon process where, in the standard case,
$\Lambda_{{\rm 2s} \rightarrow {\rm 1s}} = 8.22458 {\rm s}^{-1}$ 
\cite{goldman}.
Following Shapiro and Breit \cite{SB59}
one finds that this fundamental process has the
very steep dependence
\begin{equation}
\Lambda_{{\rm 2s} \rightarrow {\rm 1s}} \propto \alpha^8.
\end{equation}
To see how the process of recombination changes with $\alpha$ we have
plotted the evolution of the ionisation fraction $x_e$ as a function
of redshift for different values of $x \equiv 
\alpha/\alpha_0$ in Fig.~1. If $\alpha$ 
increases
interactions become stronger and equilibrium is maintained longer, meaning
that the final ionisation fraction becomes smaller. This is exactly the
trend seen in Fig.~1.
\begin{figure}[h]
\begin{center}
\epsfysize=7truecm\epsfbox{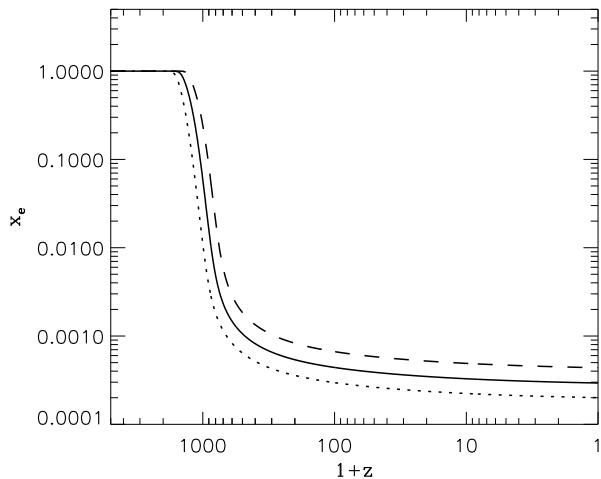}
\vspace{0truecm}
\end{center}
\baselineskip 17pt
\caption{The ionisation fraction as a function of redshift for three
different values of $x \equiv \alpha/\alpha_0$. The solid curve is for
$x=1$, the dashed for $x=0.95$ and the dotted for $x=1.05$.}

\label{fig1}
\end{figure}

The fact that a lot of the parameters entering the recombination equations
are extremely sensitive to changes
in $\alpha$ brings hope that the CMBR spectrum is equally sensitive to
changes in $\alpha$. This is exactly the case, as will be discussed in the 
next section. In Fig.~2 we have shown the CMBR fluctuation spectrum for
a standard CDM model with two different values of $x$. There are seen to
be very substantial changes, even for a quite small change in $x$.
\begin{figure}[h]
\begin{center}
\epsfysize=7truecm\epsfbox{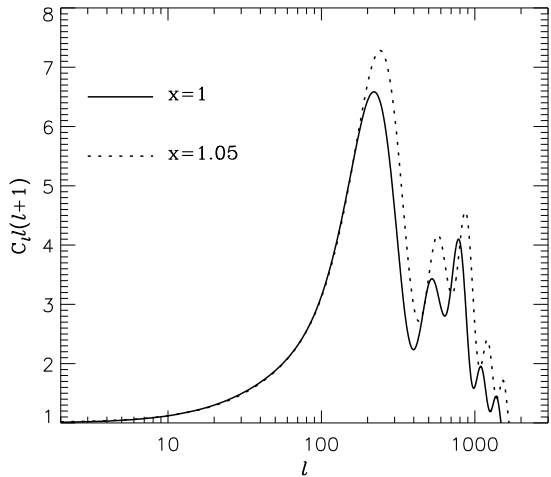}
\vspace{0truecm}
\end{center}
\baselineskip 17pt
\caption{CMBR fluctuation spectra for two different values of $x$. The
spectrum has been normalised to the quadrupole fluctuation $6C_2$.}

\label{fig2}
\end{figure}

\section{CMBR sensitivity to changes in $\alpha$}

The key question is now whether or not it will be possible to detect
deviations in $\alpha$ relative to the standard value. In order to
estimate the sensitivity of the CMBR data, we use a standard
technique for this purpose.
Since we have no usable data at present we can only provide what is
called {\it error forecasting} \cite{tegmark}. 
To do this we choose an underlying 
cosmological model (in our case standard CDM) and determine how
precisely the cosmological parameters can be determined. This method
has been described in great detail elsewhere \cite{tegmark,jungman}
 and we shall not go into
details. 
The cosmological model can be described by a vector of parameters and
in our calculations we work with the following set
\begin{equation}
\Theta = (\Omega, \Omega_b, \Lambda, h,n,N_\nu,\tau,\alpha).
\end{equation}
Here $\tau$ is the optical depth due to possible reionisation
and $n$ is the spectral index.
The standard CDM model which we choose as our reference is then given by
the vector
\begin{equation}
\Theta_{\rm CDM} = (1, 0.08, 0, 0.5,1,3,0,\alpha_0).
\end{equation}

The main point is then to calculate the so-called Fisher matrix,
which is given by
\begin{equation}
I_{ij} = \sum_{l=2}^{l_{\rm max}} (2l+1) \left[C_l + C_{l,{\rm error}}
\right]^{-2} \frac{\partial C_l}{\partial \theta_i}
\frac{\partial C_l}{\partial \theta_j},
\end{equation}
where $C_{l,{\rm error}}$ represents the experimental error. Following
Lopez {\it et al.} \cite{lopez}
we shall neglect the experimental error and only 
take into account the ``error'' induced by cosmic variance.
It can then be shown that the standard error in estimating any parameter
is of the order $\sigma_i^2 \simeq (I^{-1})_{ii}$. Specifically, if 
all parameters are allowed to vary simultaneously one obtains
\begin{equation}
\sigma_i^2 \simeq (I^{-1})_{ii},
\end{equation}
whereas if all parameters except $\theta_i$ have been determined, it is
\begin{equation}
\sigma_i^2 \simeq (I_{ii})^{-1}.
\end{equation}

Using our cosmological model as given above, we have calculated the
expected precision to which $x \equiv \alpha/\alpha_0$ can be determined.
The results of this calculation have been shown in Fig.~3 as a function
of the maximum measured $l$-value. We note here that
$l_{\rm max}({\rm MAP}) \simeq 1000$ and $l_{\rm max}({\rm PLANCK}) 
\simeq 2500$.
Fortunately, since the CMBR spectrum is very sensitive to changes in $x$,
it seems possible to detect changes as small as $10^{-3}-10^{-2}$
even if all cosmological parameters must be determined simultaneously.
To be on the conservative side we estimate that $\delta x \leq 10^{-2}$
is a realistic obtainable precision.
\begin{figure}[h]
\begin{center}
\epsfysize=7truecm\epsfbox{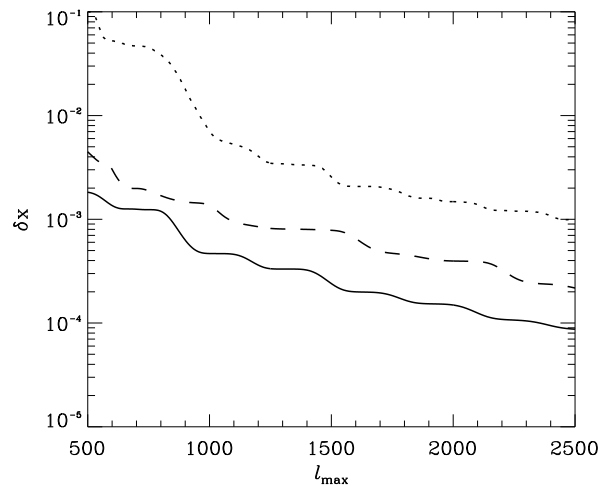}
\vspace{0truecm}
\end{center}
\baselineskip 17pt
\caption{The expected standard error $\delta x$ 
as a function of the maximum measured $l$ in a CMBR measurement of
$x$. The solid curve assumes that all other parameters are known whereas
the dotted line assumes no prior knowledge of any parameter.
The dashed curve is with $\Omega_b$ and $h$ held fixed but all other
parameters allowed to vary.}

\label{fig3}
\end{figure}

Of course in the event that all other parameters can be determined by
other means it should be possible to detect $\delta x \leq 10^{-4}$.
This is surely not within reach in the foreseeable future, but it is
still interesting to look at what other parameters it is most 
important to determine in order to obtain a better constraint on 
$\delta x$. It turns out that $x$ is most degenerate with $\Omega_b$
and $h$, and Fig.~3 we have also shown the standard error on $x$
assuming that these two parameters are held fixed at their fiducial
values $h = 0.5$ and $\Omega_b = 0.08$. If these two parameters can 
be determined by other means a factor of 3-5 improvement in the 
precision should be possible.
A possible determination of $\Omega_b$ should come from Big Bang
nucleosynthesis (BBN) arguments. Especially the new measurements
of deuterium in quasar absorption systems seem very promising in this
regard \cite{deuterium}. 
As for a measurement of the Hubble parameter perhaps the
most promising method is to use what is called {\it cosmic 
complementarity}, namely the fact that a joint use of CMBR
measurements and large scale galaxy surveys like the Sloan Digital
Sky Survey (SDSS) break some of the degeneracy in the CMBR measurements
and allows for a more precise determination of $h$ \cite{EHT98}.

\section{Discussion}

To compare with other constraints we convert the above constraint on
$\delta x$ to a 
constraint on redshift and time evolution of $\alpha$ and obtain
\begin{equation}
|\alpha^{-1} d\alpha/dz| \leq 9 \times 10^{-5},
\end{equation}
or 
\begin{equation}
|\alpha^{-1} d\alpha/dt| \leq 7 \times 10^{-13} \, {\rm y}^{-1}.
\end{equation}

This number should be compared with the constraints coming from other 
sources.
Especially constraints coming from the line shift of quasar absorption
systems have been extremely useful in providing constraints on the
time evolution of $\alpha$ \cite{savedoff,CS95,varshal96}. 
The most recent such measurement is that
of Varshalovich {\it et al.} \cite{varshal96}, who obtained
\begin{equation}
(\Delta \alpha/\alpha)_{\ z \simeq 3} \leq 1.6 \times 10^{-4}.
\end{equation}
This would correspond to 
\begin{equation}
|\alpha^{-1} d\alpha/dz| \leq 6 \times 10^{-5}
\end{equation}
or
\begin{equation}
|\alpha^{-1} d\alpha/dt| \leq 1.6 \times 10^{-14}\, {\rm y}^{-1}.
\end{equation}
It should be noted that there is actually a claim that time variation
in $\alpha$ has been detected from QSO data \cite{webb}. Here, a change of
\begin{equation}
\Delta \alpha/\alpha = -1.5 \pm 0.3 \times 10^{-5}
\end{equation}
has been reported.

It thus seems that the possible constraints on changes in $\alpha$
coming from the CMBR data will be almost as good as those from
QSO absorption systems if $\alpha$ evolves linearly in time, and potentially
better if the evolution is non-linear.
Moreover it is important to have reliable constraints from different
epochs in the evolution of our universe.

We also note that it is possible to constrain the evolution
of $\alpha$ using arguments from Big Bang nucleosynthesis. However, these
are much more model dependent than those obtainable from CMBR data.
Kolb, Perry and Walker \cite{KPW86} found an upper limit of
\begin{equation}
|\alpha^{-1} d\alpha/dt| \leq 1.5 \times 10^{-14}\, {\rm y}^{-1},
\end{equation}
but, as previously mentioned,
this is based on a specific assumption of how changes in $\alpha$
affect the neutron to proton mass ratio, an assumption which is at best
uncertain.

Finally there are also quite severe constraints coming from laboratory
experiments and other terrestrial sources. Presently, the best laboratory
limit is that of Prestage, Tjoelker and Maleki \cite{PTM95} who obtained
\begin{equation}
|\alpha^{-1} d\alpha/dt| \leq 3.7 \times 10^{-14} \, {\rm y}^{-1}.
\end{equation}
The other very interesting terrestrial constraint comes from the Oklo
natural fission reactor in Gabon 
\cite{DD96,shlyakter}. The most recent discussion is that
of Damour and Dyson \cite{DD96} who derived the limit
\begin{equation}
|\alpha^{-1} d\alpha/dt| \leq 5 \times 10^{-17} \, {\rm y}^{-1}.
\end{equation}

In conclusion, we have calculated the expected sensitivity of CMBR
measurements to changes in the electromagnetic fine structure constant 
$\alpha$. It was found that CMBR should provide a constraint which is
on the same order of magnitude as other known constraints from cosmology
and terrestrial sources. Also, this type of constraint is completely
independent of all other existing limits, a fact which makes it very 
interesting.
It should perhaps also be noted here that CMBR data can potentially also
be used to constrain the time variation of other fundamental constants
such as $m_e$, $G_F$ or $G_N$.

{\it Note added} -- After this paper had been submitted another paper
by Kaplinghat, Scherrer and Turner \cite{KST98}
on the same subject has appeared. Using the same methods they reach 
conclusions very similar to those presented in the present paper.



\begin{references}

\bibitem{superstring}See for instance M.~B.~Green and J.~H.~Schwarz and
E.~Witten, {\it Superstring theory}, Cambridge University Press (1987)
and P.~Sisterna and H.~Vucetich, Phys.\ Rev.\ D {\bf 41}, 1034 (1990).

\bibitem{PDG98}C.~Caso {\it et al.}, The European Physical Journal
{\bf C3} (1998).

\bibitem{tegmark}See for instance 
M.~Tegmark, in Proc.\ Enrico Fermi Summer School,
Course CXXXII, Varenna (1995), astro-ph/9511148.

\bibitem{cobe}G.~F.~Smoot {\it et al.}, Astrophys.\ J.\ {\bf 396},
L1 (1992).

\bibitem{MAP+PLANCK}For information on these missions see the internet
pages for MAP (http://map.gsfc.nasa.gov) and PLANCK \\
(http://astro.estec.esa.nl/Planck/).

\bibitem{hannestad}
R.~E.~Lopez {\it et al.}, astro-ph/9806116 (1998);
S.~Hannestad, Phys.\ Lett.\ {\bf B431}, 363 (1998);
S.~Hannestad and G.~Raffelt, astro-ph/9805223 (1998).

\bibitem{HET98}W.~Hu, D.~J.~Eisenstein and M.~Tegmark,
astro-ph/9806362 (1998).

\bibitem{SZ96}U.~Seljak and M.~Zaldarriaga, 
Astrophys.\ J.\ {\bf 469}, 437 (1996).

\bibitem{weinberg}S.~Weinberg, {\it The quantum theory of fields, Vol.~1},
Cambridge University Press (1995).

\bibitem{MB95}C.-P.~Ma and E.~Bertschinger, Astrohys.\ J.\ {\bf 455}, 7 
(1995).

\bibitem{peebles}P.~J.~E.~Peebles, Astrophys.\ J.\ {\bf 153}, 1 (1968).
See also B.~J.~T.~Jones and R.~F.~G.~Wyse, Astron.\ Astrophys.\
{\bf 149}, 144 (1985).

\bibitem{goldman}S.~P.~Goldman, Phys.\ Rev.\ A {\bf 40}, 1185 (1989).

\bibitem{SB59}J.~Shapiro and G.~Breit, Phys.\ Rev.\ {\bf 113}, 179 (1959).

\bibitem{jungman}G.~Jungman {\it et al.},
Phys.\ Rev.\ D {\bf 54}, 1332 (1996).

\bibitem{lopez}R.~E.~Lopez {\it et al.}, astro-ph/9803095 (1998).

\bibitem{deuterium}See for instance S.~Burles and D.~Tytler,
Space Science Rev.\ {\bf 84}, 65 (1998).

\bibitem{EHT98}D.~J.~Eisenstein, W.~Hu and M.~Tegmark,
astro-ph/9807130 (1998).

\bibitem{savedoff}M.~P.~Savedoff, Nature {\bf 178}, 688 (1956).

\bibitem{CS95}L.~L.~Cowie and A.~Songaila, Astrophys.\ J.\ {\bf 453},
596 (1995).

\bibitem{varshal96}D.~A~Varshalovich {\it et al.}, astro-ph/9607098 (1996).

\bibitem{webb}J.~K.~Webb {\it et al.}, astro-ph/9803165 (1998).

\bibitem{KPW86}E.~W.~Kolb, M.~J.~Perry and T.~P.~Walker,
Phys.\ Rev.\ D {\bf 33}, 869 (1986).

\bibitem{PTM95}J.~Prestage, R.~Tjoelker and L.~Maleki,
Phys.\ Rev.\ Lett.\ {\bf 74}, 3511 (1995).

\bibitem{DD96}T.~Damour and F.~Dyson, Nucl.\ Phys. {\bf B480}, 37 (1996).

\bibitem{shlyakter}A.~I.~Shlyakter, Nature {\bf 264}, 340 (1976).

\bibitem{KST98}M.~Kaplinghat, R.~J.~Scherrer and M.~S.~Turner,
astro-ph/9810133 (1998).

\end{references}
\end{document}